\documentclass[10pt]{article}

\usepackage{amsmath}
\usepackage{amssymb}

\usepackage{graphicx}
\usepackage{threeparttable}
\usepackage{enumerate}
\usepackage{cite}

\usepackage{color}

\usepackage[labelfont=bf,labelsep=period,justification=raggedright]{caption}

\makeatletter
\renewcommand{\@biblabel}[1]{\quad#1.}
\makeatother

\date{}

\begin{document}

\begin{flushleft}
{\Large
\textbf{Consistence beats causality in recommender systems}
}
\\
Xuzhen Zhu$^{1}$,
Hui Tian$^{1}$,
Zheng Hu$^{1}$,
Ping Zhang$^{1}$,
Tao Zhou$^{2,3}$
\\
\bf{1} State Key Laboratory of Networking and Switching Technology, Beijing University of Posts and Telecommunications, Beijing, 100876, China
\\
\bf{2} CompleX Lab, Web Sciences Center, University of Electronic Science and Technology of China, Chengdu, 610054, China
\\
\bf{3} Big Data Research Center, University of Electronic Science and Technology of China, Chengdu, 610054, China
\\

\end{flushleft}

\section*{Abstract}
The explosive growth of information challenges people's capability in finding out items fitting to their own interests. Recommender systems provide an efficient solution by automatically push possibly relevant items to users according to their past preferences. Recommendation algorithms usually embody the causality from what having been collected to what should be recommended. In this article, we argue that in many cases, a user's interests are stable, and thus the previous and future preferences are highly consistent. The temporal order of collections then does not necessarily imply a causality relationship. We further propose a consistence-based algorithm that outperforms the state-of-the-art recommendation algorithms in disparate real data sets, including \textit{Netflix}, \textit{MovieLens}, \textit{Amazon} and \textit{Rate Your Music}.


\section*{Introduction}
With the rapid development of Internet~\cite{zhang2008evolution,pastor2007evolution}, World Wide Web ~\cite{broder2000graph,doan2011crowdsourcing} and intelligent mobile phone technologies~\cite{goggin2012cell,zheng2010smart}, our social lives have been greatly changed. At the same time, we are facing inconceivably huge amount of information, such as trillions of web pages, billions of e-commerce products and millions of videos, largely challenging our information processing capability to effectively find out our interested items. Using searching queries as keywords, search engine~\cite{lawrence1999accessibility,jansen2001review,croft2010search} breaks such dilemma via powerful information retrieval, however, it strongly tends to provide users with popular information while fails to match niche items with personalized interests. In addition, it cannot dig out the things you like that are not easy to be described by a few searching queries. Under those limitations, recommender systems~\cite{resnick1997recommender} show excellent performance in providing personalized recommendations.

Due to the ever-decreasing costs of data storage and processing, recommender systems gradually spread to most areas
in our lives. Venders utilize our purchase records to recommend relevant products to enhance sales~\cite{linden2003amazon}, social web sites analyze social links to help us find more new friends~\cite{ellison2007benefits,qian2013personalized}, and online radio stations remember skipped songs to better serve us in the future~\cite{moerchen2006understandable}. In general, whenever there is plenty of diverse products and customers are not alike, personalized recommendation may help to deliver the right content to the right person. This is particularly the case for those Internet-based companies that try to make use of the so-called long-tail of products which are rarely purchased but due to their multitude they can yield considerable profits~\cite{anderson2008long}. For example, on Amazon.com, 20\% to 40\% sales come from products that do not belong to the shop's 100,000 most popular products~\cite{brynjolfsson2003consumer}. A recommender system may hence have significant impact on a company's revenues: for example, as mentioned by Sanders in the 3rd ACM Conference on Recommender Systems, 60\% of DVDs rented by Netflix are selected based on personalized recommendations. As discussed in~\cite{schafer1999recommender}, recommender systems not only help decide which products should be offer to an individual customer, they also increase cross-sell by suggesting additional products to the customers and improve consumer loyalty because consumers tend to return to the sites that best serve their needs (see~\cite{chen2004impact} for an empirical analysis on the impacts of recommendations and consumer feedback on sales at Amazon.com).

Therefore, driven by the significance in economy and society~\cite{schafer2001commerce,huang2007comparison,wei2007survey}, studies on recommender systems are progressing prosperously, and the design of an efficient recommendation algorithm attracts a wide range of interests from engineering science to marketing practice, from mathematical analysis to physics community (see the review articles~\cite{adomavicius2005toward,lu2012recommender,shapira2011recommender} and the references therein). Many recommendation techniques have been developed, including collaborative filtering~\cite{herlocker2004evaluating}, content-based analysis~\cite{ansari2000internet,pazzani2007content,adomavicius2005incorporating}, knowledge-based analysis~\cite{trewin2000knowledge}, time-aware analysis~\cite{petridou2008time,campos2014time}, tag-aware analysis~\cite{zhang2011tag,tso2008tag}, social recommendation~\cite{ma2008sorec,shepitsen2008personalized}, constraint-based analysis~\cite{felfernig2008constraint}, spectral analysis~\cite{maslov2001extracting}, iterative refinement~\cite{ren2008information}, principle component analysis~\cite{goldberg2001eigentaste}, hybrid algorithms~\cite{burke2002hybrid,zhou2010solving}, diffusion-based algorithms~\cite{zhang2007recommendation,zhou2007bipartite,zhang2008heat}, and so on. This work is closely related to the diffusion-based methods, which have already found applications in many real e-commerce systems, see for example, taobao.com and baifendian.com. Recently, the original methods get improved by considering the effects of initial resource distribution~\cite{zhou2008effect,jia2008new}, correlations biased diffusion~\cite{zhou2009accurate,liu2010personal,liu2010degree,lu2011information,liu2011information}, users' tastes~\cite{liu2009effects}, temporal effects \cite{Liu2009link}, and so on.

In general, a recommender system tries to find out users' habits and recommends uncollected objects to them based on their historical records. Most known recommendation algorithms embodies causality relationship, that is, it recommends a certain object because of some already collected objects. In such situation, temporal order is a very critical factor. Looking at an simplified example in figure 1(a), if the target user has read the textbook \emph{Algorithm}, we will prefer to recommend \emph{Data Mining} instead of \emph{Data Structure}, since the latter one should be already studied before \emph{Algorithm}. However, in many cases, such as food, music, movies, etc., such relationship does not work and the temporal order of a user's choices do not reflect any causality. As shown in figure 1(b), if the target user has watched the movie \emph{Star Trek Into Darkness}, we can infer he/she likes science fiction movies, and recommend both movies \emph{The Day After Tomorrow} and \emph{Cloud Atlas}, regardless which one should be watched before or after another one.

\begin{figure}
\centering
\includegraphics[width=16cm]{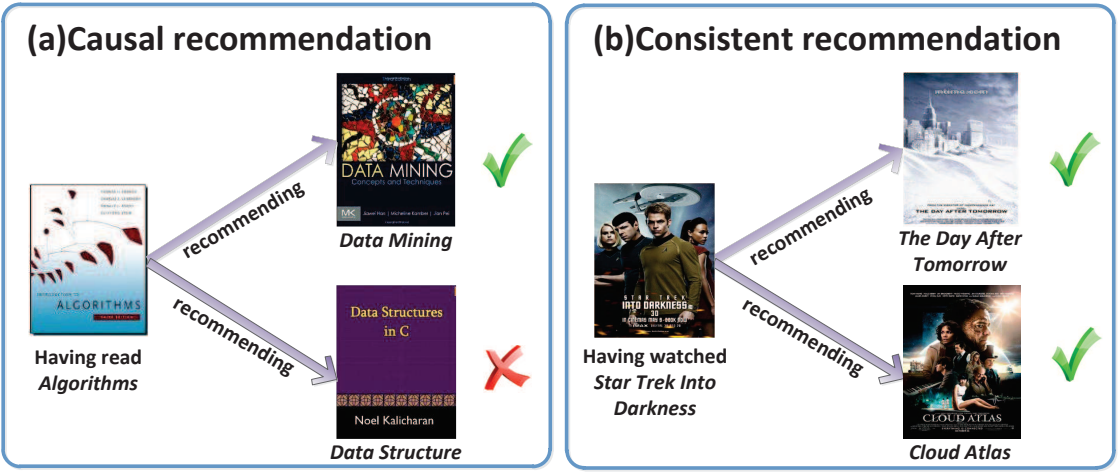}
\caption{Illustration of causal and consistent recommendation.}
\label{fig:causality_consistence}
\end{figure}

As above mentioned, some selections result from causality with temporal order, while others may only reflect consistent interests. We argue that the considerable part of selections can be explained by consistence, while many known algorithms~(e.g., the network-based inference~\cite{zhou2007bipartite}) embody the causality hypothesis: from what having been collected to what should be recommended. In this article, based on consistence, we propose a novel algorithm named consistence-based inference (CBI). We have tested our algorithm on four real datasets: \textit{MovieLens}, \textit{Netflix}, \textit{Amazon} and \textit{Rate Your Music}~(RYM). The results demonstrate higher accuracy, diversity and novelty of CBI compared with some baseline algorithms: global ranking method~(GRM), collaborative filtering~(CF), network-based inference~(NBI) and heterogenous network-based inference~(HNBI). By integrating the causality and consistence, we further propose a so-called unbalanced CBI~(UCBI) algorithm, which performs even remarkably better than CBI.

\section*{Results}
\begin{figure}
\centering
\includegraphics[width=10cm]{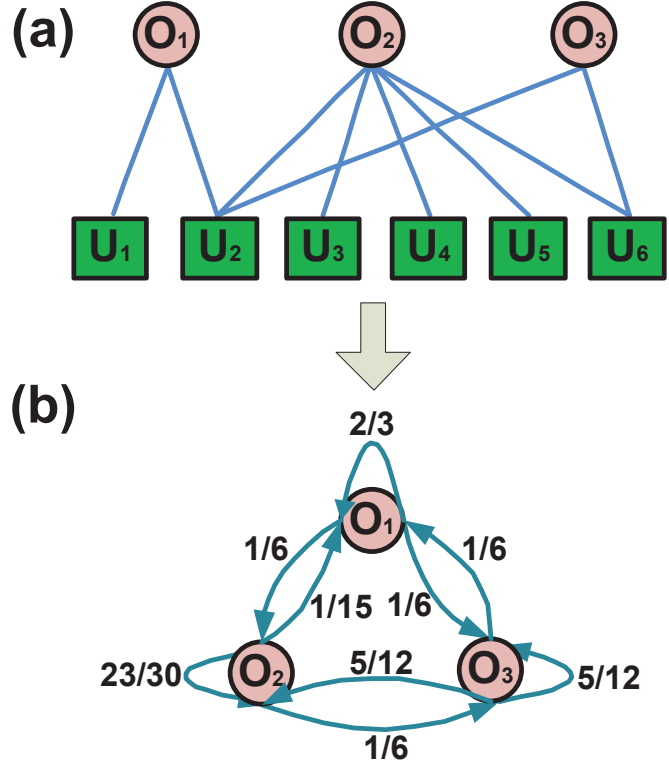}
\caption{An illustration of NBI. Subgraph (a): a bipartite network with objects denoted by rectangles and users by circles, and a user is connected with an object if this user has collected this object. Subgraph (b): the weights $w_{ij}$ (beside the arrow from node $j$ to node $i$) corresponding to the above bipartite network after the projection from user-object network to object-object network, by Eq. (\ref{equ:NBI}).}
\label{fig:NBI}
\end{figure}

A recommender system consists of users and objects, and each user has collected some objects.
Denoting the object set as $O = \{o_1, o_2, \cdots, o_n\}$, user set as $U = \{u_1, u_2, \cdots, u_m\}$ and link set as $E$, the recommender system can be fully described by an $n \times m$ adjacent matrix $A = \{a_{ij}\}$, where $a_{ij} = 1$ if $o_i$ is collected by $u_j$, and $a_{ij} = 0$ otherwise. Accordingly, we can visualize the recommender system as a bipartite network $G$ with $m+n$ nodes, where the degrees of an object $o_i$ and user $u_l$, $k(o_i)$ and $k(u_l)$, respectively represent the number of users who have collected $o_i$ and the number of objects collected by user $u_l$. Mathematically speaking, for a given user, a recommendation algorithm generates a rank for all the objects he/she has not collected yet and recommends the top-$L$ uncollected objects to this user, with $L$ denoting the length of the recommendation list. In this article, we fix $L=50$ and we have checked that the major conclusions are not sensitive to the value of $L$.

Among many known algorithms, NBI is fast, robust and relatively accurate \cite{lu2012recommender}, which is a good choice as the benchmark algorithm since it embodies the causality hypothesis. According to the standard NBI \cite{zhou2007bipartite}, given the target user $u_l$, the preference to select an object $o_i$ because of a prior selection of another object $o_j$ is defined as
\begin{equation}\label{equ:NBI}
w_{ij}=\frac{1}{k(o_j)}\sum_{l=1}^{m}\frac{a_{il}a_{jl}}{k(u_l)},
\end{equation}
which results from a simple random walk from node $o_j$ to node $o_i$. Denote by $f$ the initial collection vector of user $u_l$, where $f_j=1$ if user $u_l$ has collected object $o_j$ and $f_j=0$ otherwise, the final recommendation score of an arbitrary object $o_i$ is the simple sum of contributions $w_{ij}$, where $j$ runs over all objects $o_j$ having already been collected (i.e., $f_j=1$) by $u_l$, namely
\begin{equation}
f'=Wf,
\end{equation}
where $W=(w_{ij})_{n\times n}$ is the asymmetrical weight matrix according to Eq. (1). A simple example about how to calculate $W$ is shown in figure 2. The uncollected objects with top-$L$ values in $f'$ will be recommended to user $u_l$.

Notice that, Eq. (1) only accounts for the contribution from a prior selection $o_j$ to a possible candidate $o_i$, that is to say, to which extent we would like to recommend $o_i$ \emph{because} of the prior selection of $o_j$. This is thus a typical causality-based recommendation algorithm. Instead of causality relationship, if we consider the consistence between $o_j$ and $o_i$, we need not only account for the contribution from a prior selection $o_j$ to a posterior selection $o_i$, but also the contribution from $o_i$ to $o_j$, namely whether the co-selection $(o_i,o_j)$ reflects a consistent interest of the target user $u_l$. Therefore, corresponding to the causality relationship in Eq. (1), a consistence relationship could be
\begin{equation}\label{equ:CBI}
r_{ij}^{CBI}=w_{ij}+\frac{w_{ji}}{\sum_{j'=1}^{n}w_{j'i}}~,
\end{equation}
where the normalization factor $\sum_{j'}w_{j'i}$ is used to make sure the influences from prior selections to posterior selections and from posterior selections to prior selections are comparable, namely
\begin{equation}\label{equ:normalization}
\sum_iw_{ij}=\sum_j\left(\frac{w_{ji}}{\sum_{j'}w_{j'i}}\right)=1.
\end{equation}
Denoting the corresponding weight matrix as $R^{CBI}=(r_{ij}^{CBI})_{n\times n}$, for the target user $u_l$ and his/her initial collection vector $f$, the recommendation score $f'$ can be obtained in a similar way as Eq. (2):
\begin{equation}
f'=R^{CBI}f.
\end{equation}
Analogous to the standard NBI \cite{zhou2007bipartite}, CBI is also a parameter-free algorithm.

It is very possible that the strengths of influences from prior selections to posterior selections and from posterior selections to prior selections should be different, therefore we further introduce a unbalance consistence-based inference (UCBI), where Eq. (3) is modified as
\begin{equation}\label{equ:UCBI}
r_{ij}^{UCBI}=(w_{ij})^{\alpha}+(\frac{w_{ji}}{\sum_{j'=1}^{n}w_{j'i}})^{\beta}~,
\end{equation}
and accordingly
\begin{equation}
f'=R^{UCBI}f,
\end{equation}
where $R^{UCBI}=(r_{ij}^{UCBI})_{n \times n}$. It is not surprised that the introduction of two tunable parameters $\alpha$ and $\beta$ will improve the algorithm's accuracy comparing to the standard CBI. In addition to that, we would like see: (i) how much the performance of CBI can be further improved, and (ii) the influence from which direction is stronger.

\begin{table*}[!t]
\begin{center}
\caption{Algorithms' performance on four data sets. For each algorithm with parameters, the performance indices are obtained by optimizing corresponding parameters subject to the largest AUC value, with resolution 0.01. Values in the brackets stand for the standard deviations, and the best-performed values are emphasized by boldface. The recommendation list is fixed as $L=50$ and the number of samplings for AUC value is fixed as $n=10^6$. See Supplementary Information (SI) for results on $L=10$ and $L=100$.}
\label{tab:performances}
\setlength{\tabcolsep}{1pt}
\footnotesize{
\begin{tabular}{ccccccc}
\hline\hline
Movielens & $AUC$  & $P$ & $Recall$ & $I$ & $H$ & $\langle k\rangle$ \\
\hline
GRM& 0.8569(0.0023) & 0.0508(0.0007) & 0.3419(0.0008) & 0.4085(0.0010) & 0.3991(0.0007) & 259(0.4410) \\
CF& 0.8990(0.0020) & 0.0638(0.0011) & 0.4227(0.0009) & 0.3758(0.0008) & 0.5796(0.0016) & 242(0.3724) \\
NBI& 0.9093(0.0016) & 0.0670(0.0011) & 0.4431(0.0009) & 0.3554(0.0008) & 0.6185(0.0013) & 234(0.3925) \\
HNBI& 0.9145(0.0014) & 0.0693(0.0011) & 0.4584(0.0010) & 0.3392(0.0009) & 0.6886(0.0011) & 219(0.4725) \\
CBI& 0.9249(0.0011) &  0.0705(0.0011) & 0.4651(0.0009) & 0.3348(0.0007) & 0.6877(0.0005) & 218(0.3034) \\
UCBI& \textbf{0.9339(0.0013)} & \textbf{0.0816(0.0012)} & \textbf{0.5334(0.0007)} & \textbf{0.3067(0.0008)}&\textbf{0.8191(0.0001)}&\textbf{176(0.1270)}\\
\hline
Netflix& $AUC$ & $P$ & $Recall$ & $I$ & $H$ & $\langle k\rangle$\\
\hline
GRM& 0.8101(0.0028) &  0.0160(0.0002) & 0.0766(0.0003) & 0.3580(0.0021) & 0.1627(0.0004) & 520(1.3402)\\
CF& 0.8714(0.0021) & 0.0235(0.0003) & 0.1103(0.0004) & 0.3106(0.0009) & 0.6787(0.0010) & 423(1.2803)\\
NBI& 0.8858(0.0019) & 0.0251(0.0003) & 0.1182(0.0004) & 0.2819(0.0008) & 0.7299(0.0006) & 398(1.0763)\\
HNBI& 0.8877(0.0020) & 0.0270(0.0004) & 0.1265(0.0005) & 0.2405(0.0006) & 0.8790(0.0003) & 312(0.6855)\\
CBI& 0.9056(0.0014) & 0.0268(0.0004) & 0.1260(0.0005) & 0.2142(0.0005) & 0.8314(0.0003) & 316(0.9044)\\
UCBI& \textbf{0.9173(0.0012)} & \textbf{0.0390(0.0003)} & \textbf{0.1806(0.0001)}& \textbf{0.1683(0.0003)}&\textbf{0.9346(0.0003)}&\textbf{215(0.1430)}\\
\hline
Amazon& $AUC$ & $P$ & $Recall$ & $I$ & $H$ & $\langle k\rangle$\\
\hline
GRM& 0.6409(0.0029) & 0.0036(0.00008) & 0.0727(0.00009) & \textbf{0.0709(0.0006)} & 0.0584(0.0001) & 133(0.3) \\
CF& 0.8810(0.0017) &0.0156(0.0001) & 0.2971(0.0001) & 0.0927(0.0001) & 0.8649(0.0008) & 81(0.1938) \\
NBI& 0.8844(0.0018) &0.0161(0.0001) & 0.3050(0.0001) & 0.0899(0.0001) & 0.8619(0.0006) & 81(0.1775)\\
HNBI& 0.8844(0.0018) & 0.0163(0.0001) & 0.3079(0.0001) & 0.0896(0.0001) & 0.8652(0.0006) & 81(0.1689)\\
CBI& 0.8937(0.0018) &0.0186(0.0002) & 0.3499(0.0002) & 0.0881(0.0002) & 0.9413(0.0002) & 59(0.1088)\\
UCBI& \textbf{0.8944(0.0005)} &\textbf{0.0189(0.0001)} & \textbf{0.3548(0.0001)}& 0.0861(0.0002)&\textbf{0.9650(0.0002)}&\textbf{48(0.1800)}\\
\hline
RYM& $AUC$  & $P$ & $Recall$ & $I$ & $H$ & $\langle k\rangle$\\
\hline
GRM& 0.8786(0.0001) & 0.0034(0.00001) & 0.1153(0.00002) & 0.1334(0.0003) & 0.0701(0.00007) & 1343(0.4268)\\
CF& 0.9548(0.0001) & 0.0129(0.00003) & 0.4185(0.00003) & 0.1604(0.00006) & 0.8216(0.00001) & 1114(0.5895)\\
NBI& 0.9611(0.0001) & 0.0131(0.00006) & 0.4251(0.00005) & 0.1580(0.0001) & 0.7912(0.00008) & 1195(0.7061)\\
HNBI& 0.9644(0.0001) &0.0135(0.00005) & 0.4388(0.00005) & 0.1548(0.00008) & 0.8113(0.00001) & 1154(0.5654)\\
CBI& 0.9692(0.0001) & 0.0143(0.00004) & 0.4647(0.00003) & 0.1362(0.00005) & 0.8302(0.00002) & 1075(0.5654)\\
UCBI& \textbf{0.9704(0.0002)} & \textbf{0.0152(0.00001)} & \textbf{0.4937(0.00002)}& \textbf{0.1207(0.00001)}&\textbf{0.8739(0.00003)}&\textbf{919(0.2900)}\\
\hline\hline
\end{tabular}
}
\end{center}
\end{table*}

To evaluate the algorithmic performance, we consider six well-known metrics \cite{lu2012recommender}: AUC value ($AUC$), precision ($P$) and recall ($Recall$) for accuracy, inter-similarity ($I$) and Hamming distance ($H$) for diversity, and average degree ($\langle k\rangle$) for novelty (see details in Materials and Methods). For $I$ and $\langle k\rangle$, the lower the better, while for others the larger the better. We compare CBI and UCBI with four benchmark methods (see details in Materials and Methods): global ranking method (GRM), user-based collaborative filtering (CF), network-based inference (NBI) and heterogeneous network-based inferenc (HNBI). As shown in Table 1, for all three aspects (accuracy, diversity and novelty) and all four data sets, CBI largely outperform the four benchmark algorithms, and UCBI can further improve the performance of CBI. One can thus expect that both the click rate and user experience can be enhanced by applying CBI or UCBI. Complementary to Table 1, we plot the precision-recall curves \cite{Buckland1994,Davis2006} by varying the length of recommendation list $L$. The curve in the right upper position corresponds to higher accuracy. As shown in figure 3, for all four data sets, curves show the same order from the left lower to the right upper, namely the accuracy order is $GRM < CF < NBI < HNBI < CBI < UCBI$, supporting the results presented in Table 1.

\begin{figure}[h]
\centering
\includegraphics[width=13cm]{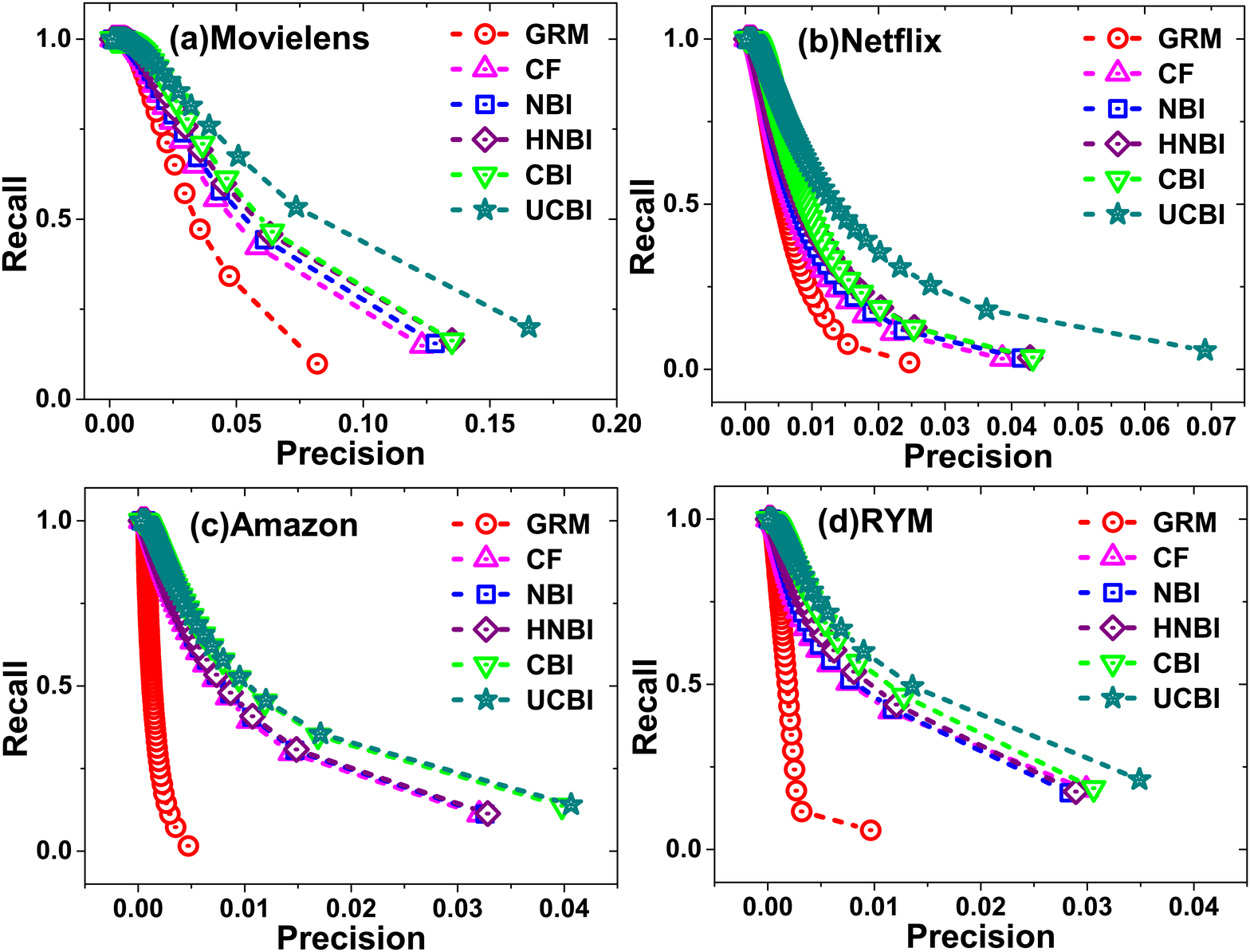}
\caption{Precision-Recall curves via varying the length of recommendation list $L$ from 1 to the cardinality of the testing set.}
\label{fig:P_R}
\end{figure}

To see the sensitivity of parameters in UCBI, figure 4 shows some representative curves by fixing $\alpha$ while varying $\beta$. Except for Amazon.com, the optimal accuracies obtained by UCBI are much higher than those by CBI. In addition, looking at the optimal values $(\alpha^*,\beta^*)$ (see figure 4(e), figure 4(f) and Table 2), for all the four data sets, $\alpha^*$ is obviously larger than $\beta^*$, suggesting that the influence from prior selections to posterior selections should be larger than the influence from posterior selections to prior selections.

\begin{figure*}[ph]
\centering
\includegraphics[width=10cm]{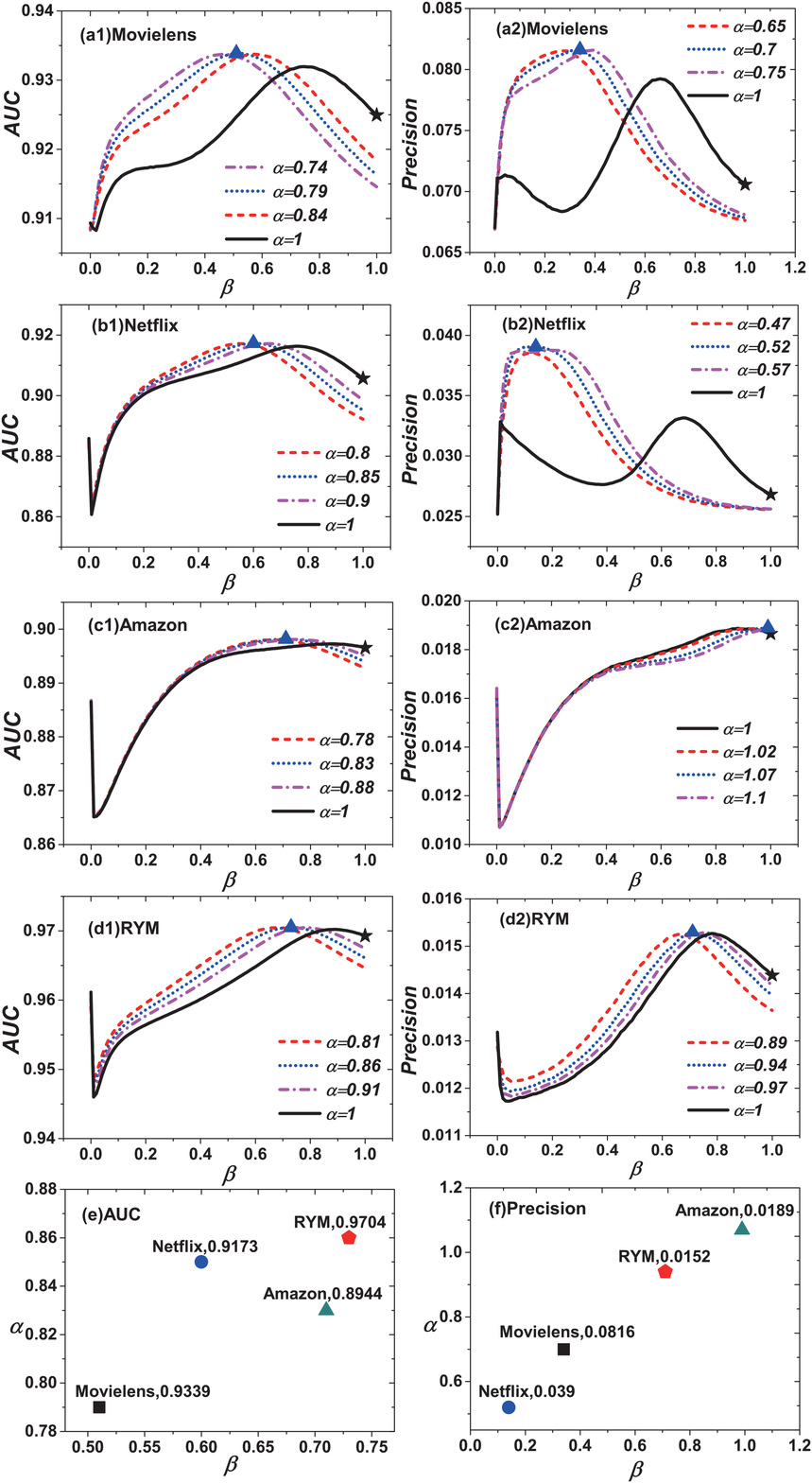}
\caption{The AUC value (left column) and precision (right column) of UCBI under different parameters $(\alpha,\beta)$ for (a) MovieLens, (b) Netflix, (c) Amazon, and (d) RYM. The optimal parameters are denoted by blue triangles and the parameters corresponding to CBI, $(\alpha,\beta)=(1,1)$, are marked by black stars. The optimal parameters for AUC and precision are directly shown in (e) and (f), respectively.}
\label{fig:precision_rank}
\end{figure*}

\begin{table}[!thp]
\caption{Optimal values of parameters $\alpha$ and $\beta$ for AUC and precision, respectively.}
\label{tab:optimalvalue}
\begin{center}
\begin{tabular}{ccccc}
\hline\hline
Data  & $\alpha^*_{AUC}$ & $\beta^*_{AUC}$ & $\alpha^*_{P}$ & $\beta^*_{P}$ \\
\hline
Movielens& 0.79 & 0.51 & 0.70 & 0.34\\
Netflix& 0.85 & 0.60 & 0.52 & 0.14\\
Amazon& 0.83 & 0.71 & 1.07 & 0.99\\
RYM& 0.86 & 0.73 & 0.94 & 0.71\\
\hline\hline
\end{tabular}
\end{center}
\end{table}

\section*{Discussions}

Recommender systems can be mathematically described in a very abstract way as a variant of link prediction problem in bipartite networks \cite{Lu2011,Shang2010}. However, the underlying decision-making processes for different kinds of recommender systems are far different to each other. For example, the click stream on free or cheap products shows different pattern from that on very expensive products, and our choices usually contain many unaware biases, such as the anchoring bias and herd behavior caused by other peers' choices and critiques \cite{Chen2012,Yang2012,Huang2012}. Therefore, the causality relationship cannot fully explain the selecting behavior of users. In a causality-based recommender system, if the target user has already selected the object $A$, and we need to choose from two candidate recommendations $B$ and $B'$, then the system will compare the recommendation strengths from $A$ to $B$ and from $A$ to $B'$. In this paper, we argue that the consistent interests play a major role in determining users' selections, hence in addition to the above operation, we should also compare the recommendation strengths from $B$ to $A$ and from $B'$ to $A$. In a word, only if recommendation strengths from $A$ to $B$ and from $B$ to $A$ are both high, we can infer that $A$ and $B$ are consistent for the target user.

According to extensive experiments on four real data sets, we show that this simple variation can remarkably improve the algorithm's accuracy, diversity and novelty. Numerical investigation suggests that the influences from $A$ to $B$ and from $B$ to $A$ are not equal, the former (aligned with the causal direction) should be stronger (as indicated by the relationship $\alpha^*>\beta^*$ in all cases). The introduction of unbalance further largely improve algorithm's performance in all three aspects. The consideration of recommendation power from unselected objects to selected objects provides a novel viewpoint to the traditional recommender systems, we hope this finding can simultaneously bring us better algorithms and more insights.

\section*{Materials and Methods}

\subsection*{Data Description}
To verify performances of recommendation algorithms, four benchmark datasets, \textit{Movielens}, \textit{Netflix}, \textit{Amazon} and \textit{Rate Your Music} (\textit{RYM}) are used, respectively\footnote{Datasets are achieved from the following web sites: http://www.grouplens.org/; http://www.netflix.com/; http://www.amazon.com/; http://rateyourmusic.com/.}.
In terms of different themes, \textit{Movielens}, \textit{Netflix} are two famous movies recommendation websites,
\textit{Amazon} is a big globalized online shopping store, selling various kinds of commodities, and \textit{RYM} is a well-known music recommendation website.
To recommend the appropriate objects, they all leverage ratings to capture users' preferences, with rating from 1 to 5 stars in \textit{Movielens, Netflix}, and \textit{Amazon} and from 1 to 10 in \textit{RYM}.
Due to a vast ocean of data information and long-tail effects, excellent algorithms are essential to their successful recommendations and can further grasp the customers' loyalty tightly in the websites.
For the sake of simplicity and privacy protection, we first anonymize the types of goods and names of users,
and then recognize preference between user and object if the ratings $\ge 3$ in \textit{Movielens}, \textit{Netflix}, \textit{Amazon} and $\ge5$ in \textit{RYM}. That is to say, only links associated with relatively high ratings are kept, which may lead to decrease of algorithm's accuracy \cite{Shang2009}. However, this issue is out of the scope of this paper. After processing, primary information of the data is summarized in Table \ref{tab:dataset}.

\begin{table}[!thp]
\caption{Primary statistics of the four data sets.}
\label{tab:dataset}
\begin{center}
\begin{tabular}{ccccc}
\hline\hline
Data  & \#Users & \#Objects & \#Links & Sparsity\\
\hline
Movielens& 943 & 1682 & 82520 & $6.3\times10^{-1}$\\
Netflix& 10000 & 6000 & 701947 & $1.17\times10^{-2}$\\
Amazon& 3604 & 4000 & 134679 & $9.24\times10^{-3}$\\
RYM& 33786 & 5381 & 613387 & $3.37\times10^{-3}$\\
\hline\hline
\end{tabular}
\end{center}
\end{table}

\subsection*{Metrics}
Before numerical experiments, the link set $E$ is randomly divided into two parts: $E^T$ is the training set consisting of 90\% links and $E^P$ is the testing set containing the rest 10\% links. The reported results are not sensitive to the size of training or testing sets, unless one of them is extremely small. Obviously, $E^P \setminus E^T=\emptyset$. The links in the testing set are regarded as unknown information and forbidden from using in training process. In the following, we introduce six performance indices for algorithms' accuracy, diversity and novelty.

\begin{enumerate}[(1)]

\item~\emph{Area Under the ROC Curve} (AUC) \cite{Hanley1982}.---
AUC attempts to measure how a recommender system can
successfully distinguish the relevant objects (those appreciated by a user) from the irrelevant objects (all the others). The
simplest way to calculate AUC is by comparing the probability that the relevant objects will be recommended with that of
the irrelevant objects. For $n$ independent comparisons (each comparison refers to choosing one relevant and one irrelevant
object), if there are $n'$ times when the relevant object has higher score than the irrelevant and $n''$ times when the scores are equal, then
\begin{equation}
AUC=\frac{n'+0.5n''}{n}~,
\end{equation}
Clearly, if all relevant objects have higher score than irrelevant objects, AUC = 1 which means a perfect recommendation
list. For a randomly ranked recommendation list, AUC = 0.5. Therefore, the degree of which AUC exceeds 0.5 indicates
the ability of a recommendation algorithm to identify relevant objects. Notice that, the sole usage of AUC may result in some misleading conclusion \cite{Lobo2008}, therefore we also consider the $L$-dependent accuracy metrics, precision and recall, and show also the precision-recall curves.

\item~\emph{Precision} ($P$) \cite{herlocker2004evaluating}.---
The number of objects recommended to a user is often limited, and even given a long recommendation
list, users usually consider only the top part of it.
For an arbitrary target user $u_i$, the precision of $u_i$, $P_i(L)$, is defined as the ratio of the number of $u_i$'s removed links $R_i(L)$ (corresponding to relevant selections), contained in the top-$L$ recommendations to $L$, say:
\begin{equation}\label{ep:Pi-L} \setlength{\abovedisplayskip}{1pt}\setlength{\belowdisplayskip}{1pt}
P_i(L) = \frac{R_i(L)}{L}~,
\end{equation}
The precision $P(L)$ of the whole system is the average of individual precisions over all users, defined as:
\begin{equation}\label{ep:Pi-L}\setlength{\abovedisplayskip}{1pt}\setlength{\belowdisplayskip}{1pt}
P(L) = \frac{1}{m}\sum_{i=1}^{m}P_i(L)~.
\end{equation}
The higher precision indicates higher accuracy.

\item~\emph{Recall} \cite{herlocker2004evaluating}.---
Recall considers the ratio of relevant selections that can be recovered in the top-$L$ recommendation list. There are two alternative ways to define recall. We can firstly define the recall of an individual user $u_i$ as
\begin{equation}\label{ep:recall}
Recall_i(L) = \frac{R_i(L)}{\left|E_i^P\right|}~,
\end{equation}
where $E_i^P$ denote the set of links associated with user $u_i$ in the testing set $E^P$ and thus $\left|E_i^P\right|$ is the number of $u_i$'s selections in the testing set. Then, similar to precision, the recall value for the whole system is defined as the average value over all users, as
\begin{equation}
Recall(L)=\frac{1}{m}\sum_{i=1}^{m}Recall_i(L)~.
\end{equation}
We can also directly define recall value as the ratio of relevant objects recovered by all the $m$ recommendation lists, as
\begin{equation}
Recall(L)=\frac{1}{\left|E^P\right|}\sum^m_{i=1}R_i(L).
\end{equation} 
In this paper, we adopt the latter metric. In addition to the separated comparisons on precision and recall, one usually plots the precision-recall curves by varying $L$ to see the overall performance \cite{Buckland1994,Davis2006}, and the curve in the right upper position indicates higher accuracy.
\\

\item~\emph{Hamming distance} ($H$) \cite{zhou2008effect}.---
The algorithm should guarantee the diversity of recommendations, viz., different users should be recommended
different objects. The intra-diversity can be quantified via the \textit{Hamming distance}. If the
overlapped number of objects in $u_i$ and $u_j$'s recommendation lists is $Q$, their Hamming distance is defined as:
\begin{equation}\label{ep:Hamming}
H_{ij} = 1 - Q/L~,
\end{equation}
Generally speaking, a more personalized recommendation list should have larger Hamming distances to other lists.
Accordingly, we use the mean value of Hamming distance,
\begin{equation}\label{ep:mean-Hamming}
H = \frac{1}{m(m-1)}\sum_{i\ne j}H_{ij}~,
\end{equation}
averaged over all user-user pairs, to measure the diversity of recommendations.
Note that, $H$ only takes into account the diversity among users.

\item~\emph{Intra-similarity} ($I$) \cite{zhou2009accurate}.---
A good algorithm should also make the recommendations to a single user diverse to some
extent \cite{ziegler2005improving}, otherwise users may feel tired for receiving many recommended objects
under the same topic. Therefore, for an arbitrary target user $u_l$,
denoting the recommended objects for $u_l$ as \{$o_1$,$o_2$,..., $o_L$\}. Using the S$\phi$ensen index \cite{sorensen1948method},
the similarity between two objects, $o_i$ and $o_j$ , can be written as:
\begin{equation}\label{ep:sij-o}
s_{ij}^{o}=\frac{1}{\sqrt{k(o_i)k(o_j)}}\sum_{l=1}^{m}a_{il}a_{jl}~.
\end{equation}
The intra-similarity of $u_l$'s recommendation list can be defined as:
\begin{equation}\label{ep:intra-simi-l}
I_l=\frac{1}{L(L-1)}\sum_{i\ne j}s_{ij}^{n}~,
\end{equation}
and the intra-similarity of the whole system is thus defined as:
\begin{equation}\label{ep:intra-simi}
I=\frac{1}{m}\sum_{l=1}^{m}I_l~.
\end{equation}
\\

\item~\emph{Average degree} ($\langle k\rangle$) \cite{zhou2008effect}.---
Given $o_{ij}$ is the j$th$ recommended object for user $u_i$, $k(o_{ij})$ represents the degree of object $o_{ij}$, so the popularity is quantified by the average degree of all recommended items:
\begin{equation}\label{ep:popularity}
\langle k\rangle=\frac{1}{mL}\sum_{i=1}^{m}\sum_{j=1}^{L}k(o_{ij})~.
\end{equation}
\end{enumerate}
The smaller $\langle k\rangle$ is preferred since to recommend niche objects usually brings better user experience.

\subsection*{Benchmark Methods}
\textbf{Global Ranking Method (GRM)}~\cite{herlocker2004evaluating}.---
GRM sorts all the objects in the descending order of degree after removing the objects that have been collected by the target user, and recommends those $L$ objects with the highest degrees.
\\
\textbf{Collaborative Filtering (CF)}~\cite{herlocker2004evaluating}.---
CF is based on measuring the similarity between users or objects. Here we consider the user-based CF, and for any two users $u_i$ and $u_j$ , their similarity is defined as the S$\phi$ensen index (for more local similarity indices as well as the comparison of them, see the Refs. \cite{liben2007link,zhou2009predicting}):
\begin{equation}\label{ep:UCF-simi}
s_{ij}=\frac{1}{\sqrt{k(u_i)k(u_j)}}\sum_{l=1}^{n}a_{li}a_{lj}~,
\end{equation}
For any user-object pair $u_i-o_j$ , if $u_i$ has not yet collected $o_j$ (i.e., $a_{ji}$ = 0), the predicted
score, $v_{ij}$ (to what extent $u_i$ likes $o_j$), is given as
\begin{equation}\label{ep:UCF-rec}
v_{ij}=\frac{\sum_{l=1,l\ne i}^{m}s_{li}a_{jl}}{\sum_{l=1,l\ne i}^{m}s_{li}}~,
\end{equation}
For any user $u_i$, all the nonzero $v_{ij}$ with $a_{ji} = 0$ are sorted in a descending order, and those objects in the top-$L$ are recommended.
\\
\textbf{Heterogenous NBI (HNBI)}~\cite{zhou2008effect}.---
HNBI is a heterogenous network-based inference algorithm dependent on the initial resource nodes' degrees, as
\begin{equation}
w_{ij}^{H}=[k(o_j)]^\beta w_{ij},
\end{equation} 
where $w_{ij}$ is defined according to Eq. (1) and other algorithmic procedures are similar to NBI.

\section*{Acknowledgments}
This work was partially supported by the National Natural Science Foundation of China (Nos. 61302077, 61433014 and 11222543), National Major
Science and Technology Special Project of China (No. 2014AA01A706), Funds for Creative Research Groups
of China (No. 61121001) and BUPT Excellent Ph.D. Students Foundation (No. CX201433). TZ acknowledges the Program for New
Century Excellent Talents in University under Grant No.NCET-11-0070 and the Special Project of Sichuan Youth
Science and Technology Innovation Research Team (Grant No. 2013TD0006).


\section*{Appendix}
\begin{table*}[hp]
\begin{center}
\caption{Analogous to Table 1, but with $L=10$.}
\label{tab:performances10}
\setlength{\tabcolsep}{1pt}
\footnotesize{
\begin{tabular}{ccccccc}
\hline\hline
Movielens & $AUC$  & $P$ & $Recall$ & $I$ & $H$ & $\langle k\rangle$ \\
\hline
GRM& 0.8570(0.0024) & 0.0818(0.0021) & 0.0987(0.0025) & 0.4585(0.0021) & 0.5113(0.0002) & 365(0.9628) \\
CF& 0.8990(0.0020) & 0.1232(0.0011) & 0.1487(0.0028) & 0.4528(0.0018) & 0.7199(0.0037) & 326(1.4314) \\
NBI& 0.9092(0.0016) & 0.1283(0.0022) & 0.1549(0.0027) & 0.4348(0.0018) & 0.7397(0.0032) & 317(1.2401) \\
HNBI& 0.9146(0.0014) & 0.1350(0.0016) & 0.1630(0.0020) & 0.4244(0.0014) & 0.7944(0.0017) & 299(1.0315) \\
CBI& 0.9249(0.0011) &  0.1350(0.0023) & 0.1630(0.0028) & 0.4278(0.0014) & 0.7739(0.0027) & 305(1.1378) \\
UCBI& \textbf{0.9339(0.0011)} & \textbf{0.1669(0.0025)} & \textbf{0.2015(0.0031)} & \textbf{0.4073(0.0018)}&\textbf{0.8754(0.0009)}&\textbf{259(1.0924)}\\
\hline
Netflix& $AUC$ & $P$ & $Recall$ & $I$ & $H$ & $\langle k\rangle$\\
\hline
GRM& 0.8102(0.0028) &  0.0246(0.0006) & 0.0204(0.0005) & 0.3941(0.0023) & 0.2256(0.0010) & 725(3.4177)\\
CF& 0.8714(0.0021) & 0.0386(0.0008) & 0.0320(0.0006) & 0.3277(0.0011) & 0.7683(0.0017) & 529(2.3665)\\
NBI& 0.8858(0.0019) & 0.0414(0.0009) & 0.0344(0.0007) & 0.2915(0.0008) & 0.8119(0.0012) & 501(2.0976)\\
HNBI& 0.8877(0.0020) & 0.0438(0.0007) & 0.0363(0.0006) & 0.2401(0.0007) & 0.9404(0.0008) & 357(2.0236)\\
CBI& 0.9056(0.0014) & 0.0431(0.0009) & 0.0358(0.0007) & 0.2065(0.0006) & 0.8969(0.0007) & 367(1.7398)\\
UCBI& \textbf{0.9173(0.0012)} & \textbf{0.0695(0.0009)} & \textbf{0.0577(0.0008)}& \textbf{0.1690(0.0004)}&\textbf{0.9657(0.0002)}&\textbf{244(0.9395)}\\
\hline
Amazon& $AUC$ & $P$ & $Recall$ & $I$ & $H$ & $\langle k\rangle$\\
\hline
GRM& 0.6409(0.0029) & 0.0047(0.0003) & 0.0162(0.0010) & \textbf{0.0911(0.0022)} & 0.0832(0.0004) & 173(0.8814) \\
CF& 0.8810(0.0017) &0.0320(0.0007) & 0.1107(0.0026) & 0.1468(0.0007) & 0.9254(0.0007) & 104(0.3520) \\
NBI& 0.8844(0.0018) &0.0325(0.0027) & 0.1125(0.0027) & 0.1427(0.0007) & 0.9193(0.0010) & 105(0.3743)\\
HNBI& 0.8844(0.0018) & 0.0328(0.0007) & 0.1135(0.0026) & 0.1423(0.0007) & 0.9214(0.0010) & 104(0.3990)\\
CBI& 0.8937(0.0018) &0.0398(0.0010) & 0.1375(0.0036) & 0.1445(0.0007) & 0.9678(0.0003) & 74(0.2561)\\
UCBI& \textbf{0.8967(0.0057)} &\textbf{0.0421(0.0031)} & \textbf{0.1454(0.0107)}& 0.1410(0.0026)&\textbf{0.9789(0.0135)}&\textbf{69(0.7100)}\\
\hline
RYM& $AUC$  & $P$ & $Recall$ & $I$ & $H$ & $\langle k\rangle$\\
\hline
GRM& 0.8786(0.0004) & 0.0096(0.00008) & 0.0585(0.0005) & 0.3274(0.0044) & 0.2142(0.0003) & 4011(2.1176)\\
CF& 0.9547(0.0004) & 0.0299(0.00007) & 0.1811(0.0004) & 0.2472(0.0003) & 0.7709(0.0002) & 2395(1.7790)\\
NBI& 0.9611(0.0001) & 0.0283(0.0003) & 0.1715(0.0007) & 0.2646(0.0003) & 0.6188(0.0001) & 3001(0.8515)\\
HNBI& 0.9644(0.0001) &0.0290(0.00009) & 0.1760(0.0005) & 0.2589(0.0003) & 0.6480(0.00001) & 2896(0.4095)\\
CBI& 0.9692(0.0002) & 0.0306(0.0001) & 0.1854(0.0010) & 0.2294(0.0004) & 0.7394(0.00006) & 2506(1.1694)\\
UCBI& \textbf{0.9705(0.0002)} & \textbf{0.0372(0.00006)} & \textbf{0.2253(0.0003)}& \textbf{0.2031(0.0002)}&\textbf{0.8433(0.0004)}&\textbf{1977(2.6425)}\\
\hline\hline
\end{tabular}
}
\end{center}
\end{table*}

\begin{table*}[hp]
\begin{center}
\caption{Analogous to Table 1, but with $L=100$.}
\label{tab:performances100}
\setlength{\tabcolsep}{1pt}
\footnotesize{
\begin{tabular}{ccccccc}
\hline\hline
Movielens & $AUC$  & $P$ & $Recall$ & $I$ & $H$ & $\langle k\rangle$ \\
\hline
GRM& 0.8570(0.0023) & 0.0372(0.0005) & 0.4497(0.0070) & 0.3601(0.0011) & 0.3377(0.0007) & 215(0.3110) \\
CF& 0.8991(0.0020) & 0.0442(0.0005) & 0.5345(0.0070) & 0.3336(0.0006) & 0.4825(0.0012) & 205(0.3753) \\
NBI& 0.9093(0.0016) & 0.0461(0.0006) & 0.5569(0.0074) & 0.3153(0.0006) & 0.5208(0.0011) & 199(0.3772) \\
HNBI& 0.9146(0.0014) & 0.0477(0.0006) & 0.5770(0.0079) & 0.3003(0.0006) & 0.5946(0.0011) & 188(0.3377) \\
CBI& 0.9249(0.0011) &  0.0486(0.0001) & 0.5877(0.0076) & 0.2868(0.0006) & 0.6248(0.0008) & 180(0.3157) \\
UCBI& \textbf{0.9339(0.0011)} & \textbf{0.0540(0.0004)} & \textbf{0.6525(0.0051)} & \textbf{0.2543(0.0006)}&\textbf{0.7795(0.0007)}&\textbf{142(0.3084)}\\
\hline
Netflix& $AUC$ & $P$ & $Recall$ & $I$ & $H$ & $\langle k\rangle$\\
\hline
GRM& 0.8101(0.0028) &  0.0136(0.0002) & 0.1130(0.0017) & 0.3395(0.0018) & 0.1416(0.0003) & 449(1.3402)\\
CF& 0.8714(0.0021) & 0.0185(0.0002) & 0.1542(0.0018) & 0.3034(0.0007) & 0.6167(0.0009) & 378(0.9545)\\
NBI& 0.8858(0.0019) & 0.0197(0.0002) & 0.1637(0.0018) & 0.2771(0.0006) & 0.6727(0.0007) & 358(0.8371)\\
HNBI& 0.8877(0.0020) & 0.0211(0.0004) & 0.1758(0.0005) & 0.2428(0.0006) & 0.8318(0.0003) & 292(0.5780)\\
CBI& 0.9056(0.0014) & 0.0210(0.0002) & 0.1743(0.0005) & 0.2167(0.0005) & 0.7861(0.0004) & 293(0.6797)\\
UCBI& \textbf{0.9173(0.0014)} & \textbf{0.0290(0.0002)} & \textbf{0.2416(0.0014)}& \textbf{0.1698(0.0001)}&\textbf{0.9117(0.0002)}&\textbf{203(0.3938)}\\
\hline
Amazon& $AUC$ & $P$ & $Recall$ & $I$ & $H$ & $\langle k\rangle$\\
\hline
GRM& 0.6409(0.0028) & 0.0030(0.00006) & 0.1045(0.0021) & \textbf{0.0601(0.0006)} & 0.0480(0.0001) & 112(0.1743) \\
CF& 0.8810(0.0017) &0.0109(0.0001) & 0.3783(0.0001) & 0.0729(0.0001) & 0.8309(0.0006) & 71(0.1036) \\
NBI& 0.8844(0.0018) &0.0112(0.0001) & 0.3888(0.0034) & 0.0705(0.0001) & 0.8287(0.0006) & 71(0.1162)\\
HNBI& 0.8927(0.0012) & 0.0126(0.00006) & 0.4356(0.0021) & 0.0644(0.0001) & 0.7771(0.0007) & 112(0.1689)\\
CBI& 0.8937(0.0018) &0.0126(0.00008) & 0.4362(0.0030) & 0.0687(0.0001) & 0.9217(0.0002) & 52(0.1088)\\
UCBI& \textbf{0.8987(0.0006)} &\textbf{0.0131(0.00007)} & \textbf{0.4529(0.0002)}& 0.0669(0.0003)&\textbf{0.9415(0.0001)}&\textbf{47(0.1244)}\\
\hline
RYM& $AUC$  & $P$ & $Recall$ & $I$ & $H$ & $\langle k\rangle$\\
\hline
GRM& 0.8786(0.0004) & 0.0027(0.00002) & 0.1653(0.0015) & 0.1199(0.0004) & 0.0520(0.00007) & 1003(0.8676)\\
CF& 0.9547(0.0002) & 0.0080(0.00002) & 0.4900(0.0011) & 0.1400(0.00009) & 0.7870(0.00008) & 841(0.2231)\\
NBI& 0.9611(0.0001) & 0.0082(0.00002) & 0.4992(0.0013) & 0.1369(0.0001) & 0.7602(0.00008) & 880(0.3522)\\
HNBI& 0.9644(0.0001) &0.0086(0.00001) & 0.5230(0.0008) & 0.1352(0.0001) & 0.7834(0.00003) & 858(0.3592)\\
CBI& 0.9692(0.0001) & 0.0091(0.00005) & 0.5519(0.00003) & 0.1186(0.0001) & 0.8195(0.00008) & 781(0.5006)\\
UCBI& \textbf{0.9705(0.0002)} & \textbf{0.0096(0.00001)} & \textbf{0.5838(0.0006)}& \textbf{0.1045(0.0001)}&\textbf{0.8681(0.00007)}&\textbf{672(0.3495)}\\
\hline\hline
\end{tabular}
}
\end{center}
\end{table*}

\end{document}